\documentclass[onecolumn,amsmath,amssymb,11pt,superscriptaddress,nofootinbib]{revtex4-2}
\usepackage[T1]{fontenc}
\usepackage{lmodern}
\usepackage[british]{babel}
\makeatletter\AtBeginDocument{\let\@elt\relax}\makeatother
\usepackage{amsmath}
\usepackage{amssymb}
\usepackage{physics}
\usepackage{amsthm}
\usepackage[]{graphicx}
\usepackage[justification=centering]{caption}
\usepackage{subcaption}
\usepackage{tensor}
\usepackage[dvipsnames]{xcolor}
\usepackage{cancel}
\usepackage{setspace}
\usepackage{fancyhdr}
\usepackage{appendix}
\usepackage{url}
\usepackage{natbib}
\usepackage{hyperref}
\hypersetup{
	colorlinks = true,
	linkbordercolor = {white},
	linkcolor=black,
	citecolor=blue,
	urlcolor=black}
\usepackage[a4paper,margin=2.5cm]{geometry}

\begin{document}
\allowdisplaybreaks
	
	\title{Quantum States Prepared by Wormholes:\\
    Long-Wavelength Deviations from Bunch-Davies}
	
	\author{George Lavrelashvili}
	\email[]{george.lavrelashvili@tsu.ge}
	\affiliation{Department of Theoretical Physics, A. Razmadze Mathematical Institute \\
		at I. Javakhishvili Tbilisi State University, GE-0193 Tbilisi, Georgia}
	\author{Jean-Luc Lehners}
	\email[]{jlehners@aei.mpg.de}
	\affiliation{Max Planck Institute for Gravitational Physics \\ (Albert Einstein Institute), 14476 Potsdam, Germany}

    
\begin{abstract}
\vspace{0.5cm}
Wineglass wormholes mediate the nucleation of baby universes out of an asymptotically Anti-de Sitter or flat spacetime. Upon materialization, the new universe naturally undergoes an inflationary phase. Here we study the quantum state of tensor and probe scalar field fluctuations that these wormhole geometries prepare, finding that they reproduce the Bunch-Davies vacuum for short-wavelength modes but lead to deviations from Bunch-Davies on large scales. These deviations, which depend on the charge of the wormhole, cause a small shift as well as oscillations in the angular power spectrum generated by an ensuing inflationary phase, and thus provide a distinctive cosmological observable. 

The wormholes are supported by either axionic or magnetic charges. In the limit of vanishing charge, they undergo a topological transition after which they contain no-boundary instantons. We find that the small-charge limit suppresses deviations from the Bunch-Davies state for both scalar and tensor fluctuations, indicating that the topological transition is smooth.
\end{abstract}

	\maketitle

\vspace{1cm}
\tableofcontents

\section{Introduction}

An intriguing idea in early universe cosmology is that our universe could have arisen via a quantum tunneling effect. Two different scenarios are commonly envisioned, either tunneling from a pre-existing spacetime, or tunneling from nothing. In either case, the hope is that such an event might be able to explain the ``initial'' conditions for the subsequent evolution of the universe. 

When tunneling from a pre-existing spacetime, there are once again at least two options: either from a dynamic spacetime, with the tunneling being mediated by Coleman-De Luccia instantons \cite{Coleman:1980aw}, or from a static space such as flat or Anti-de Sitter (AdS) space. The latter case can be mediated by Euclidean wormholes, leading to the creation of baby universes. In order for the created baby universe to expand, rather than contract, after nucleation, the mediating instanton must have a special shape: it takes the form of a (half) wormhole, with an asymptotic mouth on one side, a minimum radius in the middle, and a local maximum of the scale factor on the other side \cite{Lavrelashvili:1988un}. These wineglass-shaped wormholes, characterized by an asymptotically flat or Euclidean AdS (EAdS) mouth geometry, have recently gained significant attention \cite{Jonas:2023ipa,Jonas:2023qle,Betzios:2024oli,Lan:2024gnv,Betzios:2024zhf,Betzios:2026rbv,Lavrelashvili:2026zsw,Lavrelashvili:2026jcw,Betzios:2026fgf,Khan:2026doo}.

\begin{figure}[htbp]
    \centering
    \includegraphics[width=0.65\textwidth]{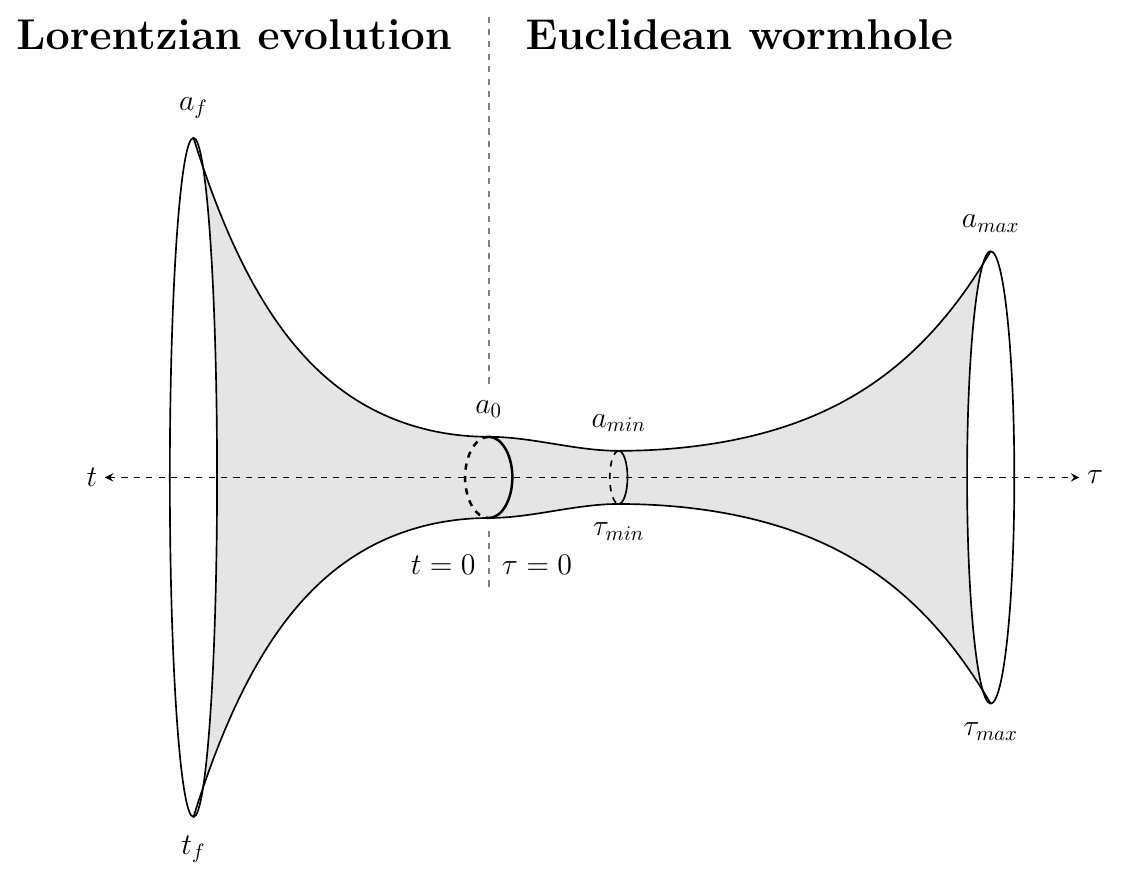}
    \includegraphics[width=0.65\textwidth]{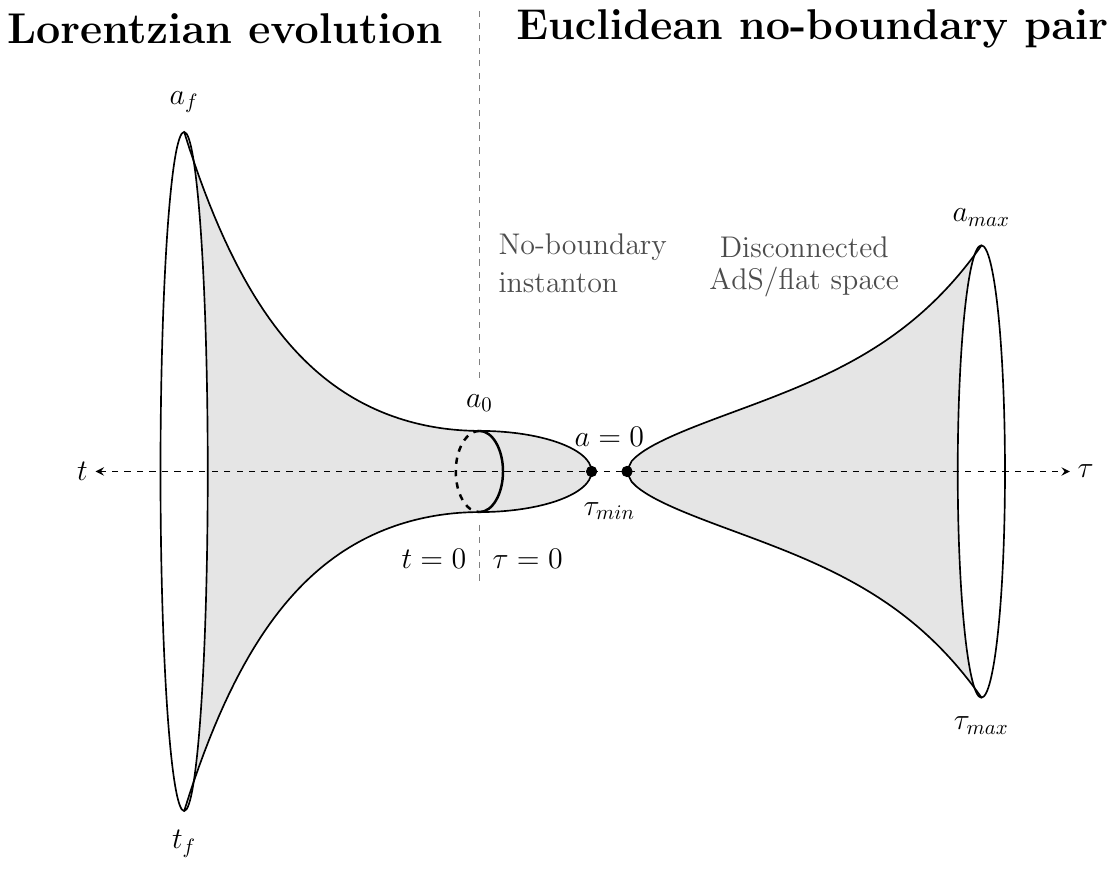}
    \caption{Wineglass wormholes and a topological transition. {\it Top:} A Euclidean wineglass wormhole mediating the nucleation of an inflationary Lorentzian universe. The asymptotic EAdS/flat region is on the right, the surface of nucleation (analytic continuation) at $\tau=0$ and the inflationary phase on the left. {\it Bottom:} In the limit that the charge vanishes, $Q \to 0$, the wormhole stem pinches off, leaving behind a Hartle-Hawking no-boundary instanton and a disconnected EAdS or flat background space.}
    \label{fig:topological_transition}
\end{figure}

Wineglass wormholes are ideally suited to lead into a phase of expansion of the universe. Then, to explain the size and flatness of the universe, as well as the perturbations seen in the cosmic microwave background (CMB), one may envision the nucleation to be followed immediately by an inflationary phase, see the top panel of Fig.~\ref{fig:topological_transition} for an illustration. In a sense then, the entire AdS-plus-wormhole (or flat-space-plus-wormhole) construction may be seen as preparing an inflationary phase of the universe. The question of interest to us here will be: which quantum state do the various fields acquire via this process?

We will look at two types of fields, namely tensor perturbations (gravitational waves) of the wormhole geometry and probe scalar fields. In inflation, it is usually assumed (often without specific reason) that the fields are in their ground state, which in the context of de Sitter (dS) or quasi-dS space is known as the Bunch-Davies vacuum \cite{Bunch:1978yq}. Here we will be interested in the question of whether or not wineglass wormholes generate this Bunch-Davies (BD) state, or a different state. What we find is that wormholes generate scale-dependent deviations from BD. For short-wavelength perturbations the BD state is indeed reproduced with great accuracy. However, interestingly, for long-wavelength perturbations there are deviations. These deviations take the form of small oscillations, combined with a shift, in the angular power spectrum of the perturbations. Moreover, the overall size of the deviations is proportional to a positive power of the charge that supports the wormhole (this can be an axionic or magnetic charge, for instance). At small values of the charges, the deviations are therefore smaller.

This result fits nicely with another recent development: in our earlier papers \cite{Lavrelashvili:2026jcw,Lavrelashvili:2026zsw} we discovered that the small-charge limit of the wineglass wormholes is highly special. As the charge is reduced, the minimum radius of the wormhole (the ``stem'' of the wineglass) becomes thinner, and in fact in the limit of zero charge it shrinks to zero size. At that point, the wineglass geometry splits into two pieces -- the base becomes pure EAdS/flat space, and the cup turns into a rounded-off dS geometry. In other words, the cup of the geometry becomes a no-boundary instanton \cite{Lehners:2023yrj}, see the bottom panel in Fig.~\ref{fig:topological_transition}. This represents a unification of wormhole and no-boundary solutions, which may therefore be regarded as different members of a common family of solutions having the axionic/magnetic charge as parameter.  

Our new results support the finding that this topological transition is smooth. We find no buildup of perturbations at the stem of the wormholes, and in fact the quantum state of both tensor and scalar perturbations reaches the BD vacuum exactly in the zero-charge limit. 

The plan of this paper is as follows: we will start by reviewing the construction of wineglass wormholes and their zero-charge no-boundary limit in section \ref{sec:wormholes}. Then, to set the scene and also in order to fix notation, we will review the calculation of cosmological perturbations in the closed slicing of dS space in section \ref{sec:dS}. Our new results regarding perturbations in the wormhole background are presented in section \ref{sec:wormholeperturbations}. And finally, we discuss our results and provide perspectives for future work in section \ref{sec:discussion}.

We work in natural units, with $8\pi G = 1, \hbar=1.$

\section{Wineglass wormholes} \label{sec:wormholes}

Two special ingredients are required for the existence of wineglass wormholes. The first is a field that can give rise to a kind of repulsive charge in Euclidean signature, such that one may obtain a spacetime tunnel that contracts and re-expands, i.e. that has the general shape of a wormhole. We will allow for two possibilities, either an axionic charge (as would be provided by the Kalb-Ramond field of string theory, for instance) or a spherically symmetric magnetic charge (as could be provided by an $SU(2)$ gauge theory, for instance). The assumption of spherical symmetry is imposed for simplicity, but is not fundamental.

The second ingredient is a scalar field $\phi$ with a non-trivial potential $V(\phi).$ This is required in order for the rim of the wormhole to correspond to a local maximum of the scale factor. We will consider potentials that contain both an extremum at zero or negative values of the potential (corresponding to the flat or EAdS asymptotic region) and another extremum at positive values of the potential (in whose vicinity inflation can occur after nucleation). 

The construction of these solutions is discussed in detail in \cite{Lavrelashvili:2026jcw}. Here we will merely provide a summary. We will work in Euclidean time $\tau,$ with metrics of the form
\begin{align}
\mathrm{d}s^2 = \mathrm{d}\tau^2 + a^2(\tau)\mathrm{d}\Omega_3^2\,,
\end{align}
where $a(\tau)$ is the scale factor and $\mathrm{d}\Omega_3^2$ denotes the line element on a unit $3$-sphere. With the ingredients listed above, the constraint and equations of motion are given by \cite{Lavrelashvili:2026zsw,Lavrelashvili:2026jcw}
\begin{align}
3 \frac{{a}^{\prime 2}}{a^2} - \frac{3}{a^2} =  \frac{1}{2} \phi^{\prime 2}  - V(\phi) - \frac{Q_a^2}{a^6} - \frac{Q_m^2}{a^4}\,, \label{constraint} \\
\phi'' + 3 \frac{{a'}}{a}\phi' - V_{,\phi} = 0 &\,, \\
 3\frac{{a''}}{a} + \phi^{\prime 2} + V - 2\frac{Q_a^2}{a^6} - \frac{Q_m^2}{a^4} = 0 & \,. \label{eoma}
\end{align}
Here primes denote derivatives w.r.t. Euclidean time $\tau,$ while later on we will denote Lorentzian time derivatives via dots. The axionic and magnetic charges are denoted by $Q_a, Q_m$ respectively, where however we will only consider one type of charge at a time.

\begin{figure}
\includegraphics[width=0.55\textwidth]{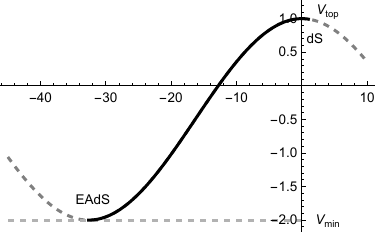}
\caption{The scalar potential contains extrema at negative and positive values. The thick curve shows the typical range covered by the scalar field in a wineglass wormhole solution.} \label{fig:potint}
\end{figure}

We will assume a scalar potential of the form
\begin{align}
V(\phi) = \frac{1-V_{min}}{2} \cos\left( \frac{\phi}{6 \sqrt{1-V_{min}}}\right) + \frac{1+V_{min}}{2}\,, \label{potential}
\end{align}
which is parametrized such that the top of the potential is at $V_{top}=1$ and the minimum at $V_{min},$ see Fig.~\ref{fig:potint}. We will consider solutions with four different values for the minimum, namely $V_{min}=-10,-1,-1/10, 0.$ The curvature of the potential at the top is chosen such that it is equal for all values of $V_{min}$ and such that it can support an (approximately) realistic inflationary phase \cite{Planck:2018jri}. This form of the potential allows the scalar field to interpolate between asymptotic EAdS or flat geometries and an inflationary region near $V_{top}.$

\begin{figure}
\includegraphics[width=0.4\textwidth]{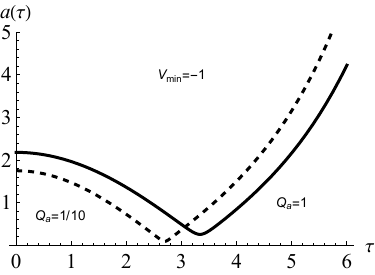}
\hspace{1cm}
\includegraphics[width=0.45\textwidth]{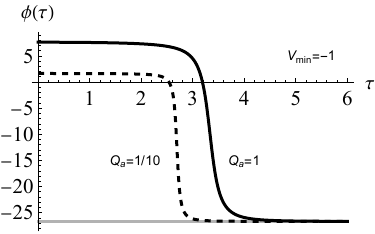}
\caption{Examples of wineglass wormholes with two different axionic charges. At smaller charge, the scalar interpolates more sharply between positive and negative potential values, and starts closer to the top of the potential. {\it Left:} Scale factor evolutions.  
{\it Right:} Scalar field evolutions.} \label{fig:numex}
\end{figure}

Wormhole solutions are found numerically, by a shooting technique in which one narrows the interval between over-shoot solutions (in which the scalar flies over the potential minimum) and under-shoot solutions (in which the scalar does not reach $V_{min}$) until the required level of precision is attained. For full details, including numerically optimized initial values, see~\cite{Lavrelashvili:2026jcw}. Two representative examples, with different values for the axionic charge $Q_a$, are shown in Fig.~\ref{fig:numex}. At smaller values of the charge, the stem of the solution becomes thinner while the scalar interpolates more sharply between positive and negative values of the potential. Also, at smaller charge, the value of the scalar field at the rim (i.e. at $\tau=0$) is closer to $V_{top},$ and in fact reaches $V_{top}$ in the limit of vanishing charge. In that limit, the scale factor approaches that of Euclidean dS space (i.e. that of a $4$-sphere) between the rim and the stem, and approaches the scale factor of EAdS (or flat, when $V_{min}=0$) space beyond the stem. Solutions with a magnetic charge behave analogously, except that they have a slightly more elongated stem region compared to the axionic solutions due to the different scaling of the charge terms in the equation of motion~\eqref{eoma}.


\section{Review of de Sitter perturbations in the closed slicing} \label{sec:dS}

When looking at perturbations in the wormhole geometry, we will be interested in possible deviations from standard inflationary perturbations. Moreover, as we have seen, wormholes with a small charge interpolate to a scalar field location very close to a local maximum of the potential. Thus the ensuing inflationary phase will be very close to pure de Sitter expansion, and in effect we will thus be interested in departures from the standard Bunch-Davies vacuum state of de Sitter spacetime \cite{Bunch:1978yq}. The calculation of Bunch-Davies mode functions is well-known in early universe cosmology, but typically it is only carried out in the flat slicing of dS, because inflation quickly disperses any spatial curvature that might be present. Here however the inflationary geometry contains a $3$-sphere, and at nucleation the sphere has a radius given by the Hubble radius, so that the spatial curvature is still very significant at first. We must therefore perform a comparison with the Bunch-Davies fluctuations in the closed slicing of dS space \cite{Gratton:2001gw,Feldbrugge:2017fcc,Lehners:2021jmv}. This is a less familiar calculation, and for this reason we will review it here.

In the closed slicing, the dS background scale factor is given by
\begin{align}
    a(t) = L \, \cosh(t/L)\,, \label{dSmetric}
\end{align}
where $L=\sqrt{3/V}$ is the radius of curvature associated with the potential energy $V.$ Note that $L$ is the minimum radius of the closed Lorentzian dS solution, reached at $t=0.$ Thus when analytically continuing our wormhole solutions with quasi-constant potential energy, we can identify $L=a_0$ at the start of the inflationary phase.

We will consider both linear tensor fluctuations and a massless scalar field whose backreaction on the geometry we will ignore (the scalar is thus a ``probe'' field). Perhaps surprisingly, in spite of the difference in helicity, these two perturbation modes give rise to the same effective action, and can thus be treated together. We start by considering transverse-traceless tensor perturbations of the $3$-dimensional spatial metric
\begin{equation}
g_{ij}(t, \mathbf{x}) = a^2(t) \left( \gamma_{ij} + h_{ij}(t, \mathbf{x}) \right),
\end{equation}
satisfying the conditions  $D^i h_{ij} = 0$ (transversality) and  $\gamma^{ij} h_{ij} = 0$ (tracelessness), and with $i,j$ taking the values $1,2,3.$ At quadratic order the action reduces to
\begin{align}
    S^{(2)}_T = \frac{1}{8}\int \mathrm{d}t\mathrm{d}x^3 \sqrt{\gamma}a^3 \left( \dot{h}_{ij} \dot{h}^{ij} - \frac{1}{a^2}D_k h_{ij} D^k h^{ij} - \frac{2}{a^2}h_{ij}h^{ij}\right)\,, \label{actiontensor}
\end{align}
where the last term arises due to the spatial curvature of the sphere and $D_k$ denotes a covariant derivative on the sphere. We can expand the perturbation in tensor hyperspherical harmonics~$T^{(n)}_{ij},$ 
\begin{align}
    h_{ij} = \sum_n h_n(t) T^{(n)}_{ij}\,,
\end{align}
where the wavenumbers take the integer values $n=3,4,5,...$. The tensor harmonics satisfy \cite{Gerlach:1978gy} $D^2 T^{(n)}_{ij} = -(n^2 - 3) T^{(n)}_{ij}.$ The equation of motion for the tensor mode amplitude $h_n(t)$ is then
\begin{equation}
\ddot{h}_n + 3\frac{\dot{a}}{a}\dot{h}_n + \frac{n^2-1}{a^2}h_n = 0\,.\label{eomperturbations}
\end{equation}

For a massless probe scalar field $\chi$, we have the action
\begin{align}
    S^{(2)}_S = \frac{1}{2}\int \mathrm{d}t\mathrm{d}x^3 \sqrt{\gamma}a^3 \left( \dot{\chi}^2 - \frac{1}{a^2}D_k \chi D^k \chi \right)\,. \label{actionscalar}
\end{align} 
This time we expand in scalar hyperspherical harmonics $S^{(n)}$ 
\begin{align}
    \chi = \sum_n h_n(t) S^{(n)}\,,
\end{align}
with wavenumbers $n=1,2,3,...$. The scalar harmonics satisfy \cite{Gerlach:1978gy} $D^2 S^{(n)} = -(n^2 - 1) S^{(n)}.$ The equation of motion for the scalar mode amplitude $h_n(t)$ is then \eqref{eomperturbations}, exactly the same as that for the tensors (because for the tensors the extra curvature terms in both the action and the harmonic decomposition cancel each other out), the only difference being the range of the wavenumbers. We can thus treat both types of perturbations simultaneously, and for this reason we used the same symbol $h_n$ for their mode functions.

The equation of motion \eqref{eomperturbations} admits two independent (not yet normalized) solutions,
\begin{align}
    f_1(t) = \left( 1 + \frac{i}{\sinh(t/L)}\right)^{\frac{n-1}{2}} \left( 1 - \frac{i}{\sinh(t/L)}\right)^{-\frac{n+1}{2}} \left( 1 - \frac{in}{\sinh(t/L)}\right)\,, \label{RealFluct1}\\ f_2(t) = \left( 1 - \frac{i}{\sinh(t/L)}\right)^{\frac{n-1}{2}} \left( 1 + \frac{i}{\sinh(t/L)}\right)^{-\frac{n+1}{2}} \left( 1 + \frac{in}{\sinh(t/L)}\right)\,, \label{RealFluct2}
\end{align}
which are complex conjugates of each other. To see which one is relevant for us, it is useful to perform an analytic continuation to Euclidean time $\tau,$ via
\begin{align}
    \frac{\tau}{L} = \frac{\pi}{2} + i \frac{t}{L}\,.
\end{align}
The scale factor then transforms to
\begin{align}
    a(\tau) = L \, \sin ( \tau/L)\,,
\end{align}
while the mode functions become
\begin{align}
    f_1(\tau) &=\frac{(1+\cos(\tau/L))^{(n-1)/2}(\cos(\tau/L)-n)}{(1-\cos(\tau/L))^{(n+1)/2}}\,, \label{dSclosedsol1}\\ f_2(\tau) &=\frac{(1-\cos(\tau/L))^{(n-1)/2}(\cos(\tau/L)+n)}{(1+\cos(\tau/L))^{(n+1)/2}}\,. \label{dSclosedsol2}
\end{align}
The scale factor corresponds to that of a $4-$sphere, and the range $0 \leq \tau \leq \pi L/2$ covers half of it, from the South Pole at $\tau=0$ to the equator at $\tau=\pi L/2,$ where the analytic continuation takes place. By direct inspection we can see that $f_1$ blows up at $\tau=0$ and must thus be discarded, while $f_2$ vanishes at the South Pole and corresponds to the regular solution. A regular dS perturbation with final value $h_f$ at time $t_f$ thus corresponds to the solution
\begin{align}
    h_n(t) = \frac{f_2(t)}{f_2(t_f)} h_f\,. \label{normBDmode}
\end{align}
One may regard the Euclidean part of the solution as preparing the Bunch-Davies vacuum state. Requiring regularity of the solution then corresponds to choosing the ``correct'' linear combination of $f_1$ and $f_2.$

We can also evaluate the action on this solution, and thus obtain the saddle-point wave function of the perturbations. The action for $h_n$ follows directly from either \eqref{actiontensor} or \eqref{actionscalar}, and its value is easy to obtain after substituting in the equation of motion \eqref{eomperturbations}, 
\begin{align}
    S^{(2)} & = \int \mathrm{d}t\mathrm{d}^3x\, \frac{1}{2} a^3 \left( \dot h_n^2 - \frac{n^2-1}{a^2}h_n^2 \right)\,, \label{actionperts}\\ S^{\textrm{on-shell}} & = 2\pi^2 \int  \mathrm{d}t \, \frac{1}{2} a^3 \left(\dot h_n^2 + h_n\ddot h_n + 3 \frac{\dot{a}}{a}\dot h_n \right) \nonumber \\ & = \pi^2 a^3 h_n \dot h_n \mid^{t=t_f} \\  & = \pi^2 a_f^3 \frac{\dot{f}_2(t_f)}{f_2(t_f)} h_f^2  \label{pertonshell}\,.
\end{align}
As we can see, the on-shell action reduces to a surface term evaluated at the final time $t_f$ (the contribution at the initial time $\tau=0$ vanishes). We can work it out more explicitly,
\begin{align}
    S^{\textrm{on-shell}}(h_f) &  = i\pi^2 a_f^2\frac{n^2-1}{n-i\sqrt{a_f^2/L^2 -1}}h_f^2 \label{pertactionexact}\\ & \approx -\pi^2 (n^2-1)L a_f h_f^2 + i \pi^2(n^3-n)L^2 h_f^2 + {\cal O}(a_f^{-1})\,, \label{pertactionlargea}
\end{align}
where we expanded in the final scale factor value $a_f=a(t_f)$ in the last line. Since the saddle-point wave function is given by $\Psi=e^{iS},$ we can interpret the first term above as a fast-growing phase, while the second gives the weighting and thus the relative probability. In other words, this is a wave function of semi-classical form, describing the production of scale-invariant perturbations out of the quantum vacuum, and with Gaussian probability distribution
\begin{align}
    |\Psi_{\textrm{perturbations}}| \approx e^{ - \frac{3\pi^2 (n^3-n)}{V}h_f^2}\,. \label{weightingnoboundary}
\end{align}
for each wavenumber $n.$

The structure of the on-shell action motivates the definition of a Riccati variable $R_n$ and the kernel $\Omega_n,$ defined via
\begin{align}
    R_n(t) \equiv -i \frac{\dot{h}_n}{h_n}\,, \qquad \Omega_n(t) \equiv a^3 R_n\,. 
\end{align}
The usefulness of these definitions lies in the fact that $R_n$ satisfies its own first-order (though non-linear) equation of motion, equivalent to \eqref{eomperturbations},
\begin{align}
    i \dot{R}_n + 3i\frac{\dot a}{a} R_n - R_n^2 + \frac{n^2-1}{a^2} = 0\,, \label{Riccatiequation}
\end{align}
while the kernel specifies the spectrum of the perturbations according to
\begin{align}
    \Psi(h_f) \propto  e^{-\frac{1}{2} V_{S^3} \Omega_n h_f^2} \quad \rightarrow \quad \langle h_f^2 \rangle = \frac{1}{2 V_{S^3}\mathrm{Re}[\Omega_n]}\,, \label{perturbativewavefunction}
\end{align}
where a factor of $V_{S^3} = 2\pi^2$ arises from the volume of the spatial $3-$sphere. The variance $\Delta_n^2,$ which may be seen as a dimensionless spectrum, can then be defined as
\begin{align}
    \Delta_n^2 \equiv (n^3-n)\langle h_f^2 \rangle\,,
\end{align}
where we use $n^3-n=(n-1)n(n+1)$ rather than simply $n^3$ in the definition, because we are dealing with the closed slicing. In the pure dS case, one obtains the late-time/large-scale-factor result
\begin{align}
    \Delta_{n,BD}^2 = \frac{1}{4\pi^2L^2} = \frac{V}{12\pi^2}\,, 
\end{align}
which we will use as reference when looking at perturbations in the wormhole geometries. The late-time Bunch-Davies variance is independent of $n,$ which is a reflection of the exact scale invariance of the spectrum.

\section{Wormhole perturbations and deviations from Bunch-Davies} \label{sec:wormholeperturbations}

We will now look at perturbations in the wormhole geometries, again by considering tensor fluctuations and a probe scalar field. Their action is given by \eqref{actionperts}, with equation of motion \eqref{eomperturbations}. The wormholes are Euclidean solutions, and can be viewed as preparing a quantum state for subsequent Lorentzian evolution -- the wormholes are not just mediating the nucleation of a baby universe, they are also preparing the ``initial'' states for all fields in the inflationary phase that follows the emergence of the baby universe. We will treat the asymptotically AdS and flat cases in turn.

\subsection{Asymptotically AdS wormholes}

When the scalar field supporting the wormhole background asymptotically reaches a negative potential value, the geometry tends to Euclidean AdS space, with scale factor
\begin{align}
    a(\tau) = {\cal L} \sinh\left[(\tau-\tau_{min})/{\cal L}\right]\,, 
\end{align}
where ${\cal L}=\sqrt{-3/V_{min}}$ is the AdS radius of curvature and where $\tau_{min}$ can be matched to the wormhole solution ($\tau_{min}$ approximately corresponds to the location of the stem). Expanding the equation of motion \eqref{eomperturbations} at large $\tau$ and choosing the decaying solution, one finds the asymptotic (unnormalized) expression
\begin{align}
    h_n(\tau) = e^{-3(\tau-\tau_{min})/{\cal L}} \, \left[ 1+ \frac{(2n^2+7)}{5} e^{-2(\tau-\tau_{min})/{\cal L}} + {\cal O}(e^{-4(\tau-\tau_{min})/{\cal L}})\right]\,. \label{AdSboundary}
\end{align}
Note that the normalization is unimportant at this stage, as it drops out of the Riccati variable $R_n = h'(\tau)/h(\tau)$ that determines the kernel. Further note that in the Euclidean region, both the Riccati variable and consequently the kernel $\Omega_n$ are real valued, and asymptotically we have
\begin{align}
    R_n(\tau) = -\frac{3}{{\cal L}}  - \frac{(4n^2+14)}{5{\cal L}} e^{-2(\tau-\tau_{min})/{\cal L}} + {\cal O}(e^{-4(\tau-\tau_{min})/{\cal L}})\,. \label{AdSboundaryRic}
\end{align}

Our strategy then is to numerically evolve the mode functions from the asymptotic EAdS region, with boundary conditions \eqref{AdSboundary} specified at some large Euclidean time $\tau_{max},$ to the rim of the wormhole at $\tau=0,$ where the analytic continuation to Lorentzian time takes place. We are looking out for potential deviations at $\tau=0$ from the Bunch-Davies vacuum that a pure de Sitter instanton would have prepared (and that is commonly assumed in inflationary models).

\begin{figure}
\includegraphics[width=0.45\textwidth]{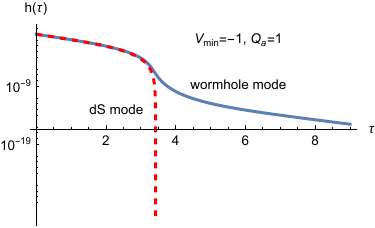}
\hspace{1cm}
\includegraphics[width=0.45\textwidth]{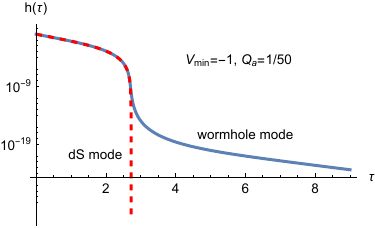}
\caption{Mode functions prepared by two different Euclidean evolutions, either via wormhole (blue solid curves) or via a dS instanton (dashed red curve). Here we use a scalar potential with $V_{min}=-1$ and show modes with wavenumber $n=5.$ {\it Left:} Large axion charge $Q_a=1$.  
{\it Right:} Small axion charge $Q_a=1/50$.} \label{fig:modefcts}
\end{figure}

We have studied examples of both the axionic and magnetic wormholes described in section~\ref{sec:wormholes}. Representative examples for wormholes with axionic charge are shown in Fig.~\ref{fig:modefcts}. What we find is that the wormhole mode functions grow very significantly throughout the entire wormhole region, and especially around the stem. For comparison, we plot a Bunch-Davies mode \eqref{normBDmode} with the same wavenumber (here $n=5$). In order for the comparison to be meaningful, the dS equator is chosen to be equal to the size of the rim of the wormhole, i.e. we fix $L=a_0.$ Moreover, the wormhole and BD modes are normalized such that they reach unity at $\tau=0.$ One can see that in the cup region of the wormhole ($\tau \lessapprox 3$), the wormhole mode function starts to (almost) coincide with the BD mode. Of course, the BD mode starts from zero at a finite location in Euclidean time, but nevertheless the two modes start to overlap rapidly. 

\begin{figure}
\includegraphics[width=0.45\textwidth]{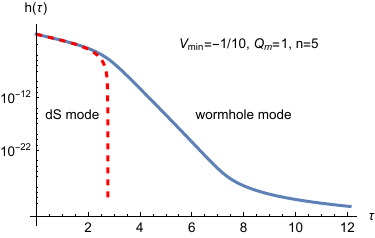}
\caption{Same as Fig.~\ref{fig:modefcts}, but this time the wormhole has magnetic charge $Q_m=1$ and the potential minimum resides at $V_{min}=-1/10.$ The magnetic wormhole has a longer stem region, which is reflected in the longer region of steep growth around $3 \lessapprox \tau \lessapprox 7.$} \label{fig:modefctsmagnetic}
\end{figure}

An analogous graph is shown in Fig.~\ref{fig:modefctsmagnetic} for the case of a magnetic wormhole. The result is almost identical, the only notable difference being the more elongated stem region during which the wormhole mode grows especially fast. But once again, at small $\tau$ the wormhole and BD modes start to overlap. Thus, by eye we can already see that wineglass wormholes produce cosmological perturbations that are qualitatively similar to the BD vacuum. 

But of course we should analyze this property more quantitatively. The physical characteristics of the perturbation modes are encoded in their wave function \eqref{perturbativewavefunction}, which is determined by the kernel $\Omega_n.$ This prompts us to analyze the difference between the wormhole kernel and the BD kernel at the surface of analytic continuation $\tau=0,$ i.e. at the onset of the ensuing inflationary phase. More precisely, we are looking at the relative difference
\begin{align}
    \frac{\Delta \Omega_n}{\Omega_n} \equiv \frac{\Omega_n - \Omega_{n,BD}}{\Omega_n} \mid_{\tau=0} = \frac{\Omega_n(0) - a_0^2\frac{n^2-1}{n}}{\Omega_n(0)}\,, \label{kerneldeviationdef}
\end{align}
where the expression for the BD kernel $\Omega_{n,BD}(\tau=0)=a_0^2\frac{n^2-1}{n}$ follows from \eqref{pertactionexact} when setting the scale factor value equal to the size of the wormhole rim, $a_f=a_0=L,$ i.e. when the ``final'' scale factor is brought all the way back to the Lorentzian origin at $t=0.$ 

\begin{figure}
\includegraphics[width=0.45\textwidth]{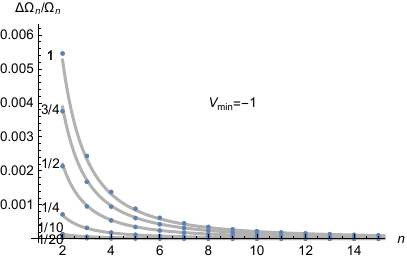}
\hspace{1cm}
\includegraphics[width=0.45\textwidth]{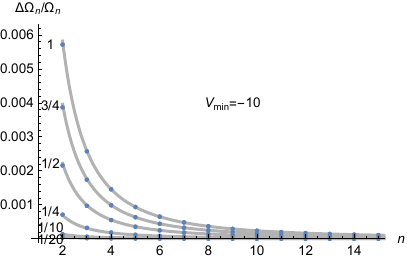}
\caption{Plots of the relative deviation of the wormhole kernel from the Bunch-Davies kernel at the surface of analytic continuation $\tau=0,$ i.e. at the moment when the ensuing inflationary phase starts. The gray curves are fits to the numerical data (blue dots). The wormhole charges, ranging from $Q_a=1/20$ to $Q_a=1$, label the respective curves. {\it Left:} $V_{min}=-1$ and axionic charges. {\it Right:} $V_{min}=-10$ and axionic charges.} \label{fig:deviation1}
\end{figure}

\begin{figure}
\includegraphics[width=0.45\textwidth]{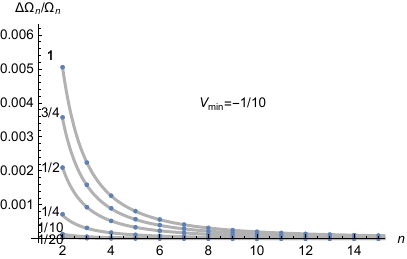}
\hspace{1cm}
\includegraphics[width=0.45\textwidth]{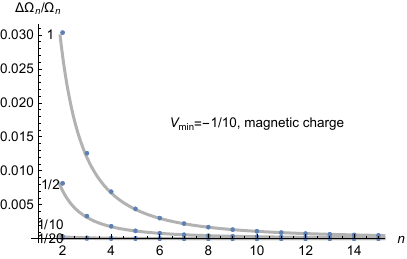}
\caption{Same as Fig.~\ref{fig:deviation1}, for $V_{min}=-1/10$.  {\it Left:} axionic charges. {\it Right:} magnetic charges.} \label{fig:deviation2}
\end{figure}

We have plotted the relative deviations of the wormhole kernel in Figs.~\ref{fig:deviation1} and \ref{fig:deviation2}, for different potential depths and various charges (the first three graphs are for axionic wormholes, and the last one for magnetic wormholes). In each case we look at wavenumbers in the range $2 \leq n \leq 15.$ Three observations are immediately apparent: 1. The deviation from the BD vacuum is largest at small wavenumbers (i.e. for long-wavelength modes) and rapidly decays at large wavenumbers. 2. The results change very little when the depth of the potential is varied. 3. The overall relative deviation is quite small, reaching a few per mille at small wavenumbers in the axionic examples, and at most a few percent in the magnetic case. 

In fact, one can fit the numerical results with rather simple functions. In the axionic case, we find the fitting function
\begin{align}
    \frac{\Delta \Omega_n}{\Omega_n} \approx \frac{Q_a^{1.8}}{10a_0^2n^2}\,, \label{axionicfit}
\end{align}
while the magnetic case is well matched by the expression
\begin{align}
    \frac{\Delta \Omega_n}{\Omega_n} \approx \frac{Q_m^{2}}{3a_0^2n^2}\,. \label{magneticfit}
\end{align}
In both cases the relative deviation scales as $1/n^2.$ On small scales, this is in fact the expected behavior on quite general grounds\footnote{We thank an anonymous referee for pointing this out to us.}. To understand this point, we may go back to the Riccati equation \eqref{Riccatiequation} and expand it at large wave number $n.$ Then one finds
\begin{align}
    R_n = \frac{n}{a} + i \, \frac{\dot{a}}{a} - \frac{1}{2n}\left(\ddot{a} + \frac{\dot{a}^2}{a} + \frac{1}{a} \right) + {\cal{O}}(n^{-2})\,.
\end{align}
Since we are comparing wormhole perturbations with de Sitter perturbations on a surface where the scale factor is equal, and where the Hubble rate vanishes in both background geometries, the only change occurs due to the difference in acceleration/curvature,
\begin{align}
    \Delta R_n = -\frac{\Delta \ddot{a}}{2n} + {\cal{O}}(n^{-2})\,.
\end{align}
Again, since in both geometries the scale factor is equal to $a_0,$ this change translates into a small-scale behavior of the kernel scaling as $1/n^2,$
\begin{align}
    \frac{\Delta \Omega_n}{\Omega_n} = \frac{\Delta R_n}{R_n} = -\frac{\Delta \ddot{a}}{2a} \frac{a^2}{n^2} + {\cal{O}}(n^{-3})\,.
\end{align}
What is special about the wormhole geometries is that the scaling of the kernel as $1/n^2$ persists to small $n,$ as the fitting formulae \eqref{axionicfit}-\eqref{magneticfit} and Figs.~\ref{fig:deviation1} and \ref{fig:deviation2} demonstrate. Together with the dependence of the relative deviation on the charges, this provides a distinctive signature of the wormholes. It is of course natural that the deviation should be larger on large scales (small $n$), since these are the scales where the wormhole geometry is most visible. On short scales, the spacetime curvature is less noticeable, and thus such modes are less sensitive to the wormhole.

The deviation from the BD vacuum may usefully be regarded as a Bogoliubov transformation. To this end, we expand the wormhole mode functions $h_n$ in a basis of BD modes, according to
\begin{align}
    h_n = \alpha_n u_n + \beta_n u_n^\star\,,
\end{align}
where $|\alpha_n|^2 - |\beta_n|^2 = 1.$ When $\beta_n \neq 0$ we have a mixing of modes, meaning that the wormhole mode functions consist of a mixture of both positive- and negative-frequency BD modes. In the definition above, we introduced the normalized BD mode $u_n,$ which is proportional to $f_2$ in \eqref{RealFluct2},
\begin{align}
    u_n(t) = \frac{1}{L\sqrt{2i(n^3-n)}}f_2(t)\,, \qquad a^3 \left( u_n \dot{u}_n^{\star} - u_n^\star \dot{u}_n\right) = i\,. \label{BDmodenorm}
\end{align}
The condition on the right determines the normalization of the modes in terms of the Wronskian. Then, in terms of the wormhole kernel $\Omega_n = -i a^3 \dot{h}_n/h_n,$ we can obtain an expression for the Bogoliubov coefficients
\begin{align}
    \frac{\beta_n}{\alpha_n} = - \frac{\Omega_n u_n + i a^3 \dot{u}_n}{\Omega_n u_n^\star + i a^3 \dot{u}_n^{\star}}\,.
\end{align}
For small $\beta_n,$ we can approximate $\alpha_n \approx 1,$ and making use of both relations in~\eqref{BDmodenorm} above as well as $\Omega_{n,BD} = -i a^3 \dot{u}_{n}/u_{n},$ we obtain the approximate relation
\begin{align}
    \beta_n \approx - \frac{\Delta \Omega_n \, u_n}{\Omega_{n,BD} u_n^\star + i a^3 \dot{u}_n^{\star}} \approx \Delta\Omega_n \, u_n^2 \propto \frac{1}{n^2}\,,
\end{align}
where we have evaluated the BD mode at the end of the wormhole/beginning of inflation. Thus we see that the wormhole generates a small admixture of negative-frequency BD modes.

\begin{figure}
\includegraphics[width=0.65\textwidth]{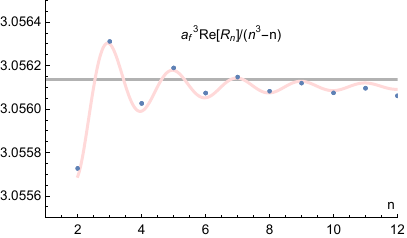}
\caption{The late-time, super-horizon kernel generated by an inflationary phase in a baby universe that nucleated via a wineglass wormhole (with $V_{min}=-1, Q_a=1/10$). The numerical results are indicated by blue dots, while the gray line indicates the reference BD result. The numerical results indicate a small oscillation in the amplitude, with an envelope decaying as $1/n^2.$ The deviations can be significantly larger when the charge of the wormhole is increased.} \label{fig:oscillations}
\end{figure}

It is of interest to evolve the perturbations to late, super-horizon scales, to see in more detail what kind of observational consequences the wormhole leads to. This can be achieved by evolving the Riccati equation \eqref{Riccatiequation} forward in (Lorentzian) time, with the numerically determined ``initial'' conditions that we have just discussed. For wormholes supported by a small charge, and which interpolate to near the top of the potential, we may approximate the inflationary phase with a dS metric \eqref{dSmetric}, with radius of curvature determined by the wormhole rim, $L=a_0.$ 

The resulting late-time kernel is plotted in Fig.~\ref{fig:oscillations}, with the numerical results shown by blue dots and the reference BD kernel indicated by the gray line. We can notice a slight overall shift of the kernel, with oscillations being superimposed. In fact, the oscillations are only significant on the largest scales and they decay with an envelope that scales as $1/n^2.$ The oscillations can be understood as arising from the interference of positive- and negative-frequency modes as the state is evolved in time.

We may also understand the oscillations directly from the Riccati equation. If we parameterize the Riccati variable in terms of a deviation $\Delta R_n \equiv R_n - R_{n,BD}$ from the BD reference case, then
we obtain (at linear order in $\Delta R_n$)
\begin{align}
    i \frac{\mathrm{d}(\Delta R_n)}{\mathrm{d}t} + (3i\frac{\dot a}{a}-2R_{n,BD}) \Delta R_n = 0\,, \label{Riccatiequationlinear}
\end{align}
where 
\begin{align}
    R_{n,BD} = \frac{n^2-1}{L\cosh(t/L)[n-i\sinh(t/L)]}\,.
\end{align}
Thus we can see that the coefficients in \eqref{Riccatiequationlinear} are complex, with evolving phases. Since the kernel only depends on the real part $\mathrm{Re}[R_n],$ and keeping in mind that modes with different wavenumbers exit the horizon at different times, it is then clear that oscillations will be seen. Moreover, it is also evident from the structure of the equation that the solution is exponential, and that consequently the original $1/n^2$ scaling of $\Delta R_n$ will be preserved. This is in good agreement with the numerics.

\subsection{Asymptotically flat wormholes}

\begin{figure}
\includegraphics[width=0.45\textwidth]{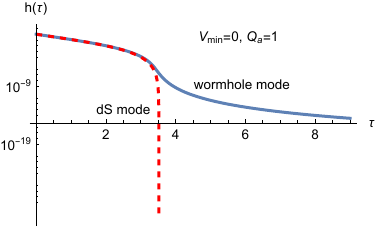}
\hspace{1cm}
\includegraphics[width=0.45\textwidth]{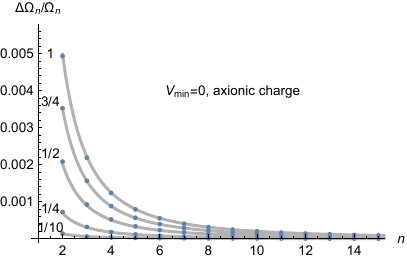}
\caption{Perturbations in asymptotically flat wormhole geometries. {\it Left:} Comparison of mode functions, with the example of a wormhole with $V_{min}=0, Q_a=1$ and wavenumber $n=5.$ The wormhole mode function is shown in blue, and for comparison we also plot an analogous dS mode with the dashed red line.
{\it Right:} The kernel of wormhole perturbations evaluated at the rim of the wormhole, $\tau=0.$ The various axion charges are indicated next to the fitting curves.} \label{fig:deviationflat}
\end{figure}

We can repeat the same calculations for wormholes that asymptote to flat space, i.e. for those with $V_{min}=0.$ Such wormholes were studied in detail in \cite{Jonas:2023ipa,Lavrelashvili:2026jcw}. In that case, we have to modify the asymptotic behavior of the mode functions. Asymptotically, the scale factor behaves as
\begin{align}
    a(\tau) \approx \tau - \tau_{min}\,,
\end{align}
where $\tau_{min}$ is again a constant shift that is located near the stem of the wormhole. Then \eqref{eomperturbations} can be expanded at large $\tau$ to yield the simple solution
\begin{align}
    h_n(\tau) = (\tau-\tau_{min})^{-1-n} \left[1+ {\cal O}((\tau-\tau_{min})^{-4})\right]\,, \qquad R_n(\tau) = -\frac{n+1}{\tau-\tau_{min}}+ {\cal O}((\tau-\tau_{min})^{-5})\,,
\end{align}
where we discarded a growing mode and retained only the decaying branch. With these initial conditions, set at a large Euclidean time $\tau=\tau_{max},$ we can then numerically evolve the mode inwards, up to the rim of the wormhole. The left graph in Fig.~\ref{fig:deviationflat} shows a representative example in blue. In much the same way as for AdS boundary conditions, we can see that the mode is amplified considerably, with the steepest increase occurring in the region of the stem. For the purpose of comparison, we also plot a dS mode that one would obtain in a Euclidean dS ($4$-sphere) geometry reaching the same equator size as the rim of the wormhole. As we can see, the two modes start overlapping more and more, indicating that the wormhole prepares a state that is close to the BD vacuum, even though the modes started out in pure flat space asymptotically.

The right graph in Fig.~\ref{fig:deviationflat} shows this more quantitatively. Here we once again plot the relative deviation of the kernel at $\tau=0,$ defined in \eqref{kerneldeviationdef}. As the graph shows, the deviations are again small, but most pronounced at small wavenumbers. In fact, they scale in pretty much exactly the same way as in the asymptotically AdS case, and again Eqs. \eqref{axionicfit} and \eqref{magneticfit} provide good fitting functions. This is perhaps not so surprising if we recall that in the AdS case, the results were only very weakly dependent on the depth of the potential. Here also, we thus find that the kernel contains a small shift scaling as $1/n^2,$ and we would expect the same kind of oscillations in the late-time power spectrum as those described in relation to Fig.~\ref{fig:oscillations}.


\section{Discussion} \label{sec:discussion}

We have studied Euclidean wormholes that mediate the nucleation of baby universes with an expanding geometry. Such wormhole solutions are of interest, as they are candidate solutions for describing what happened in the early stages of evolution of our own universe. The evolution in Euclidean time not only describes the emergence of the background geometry, but it also prepares the quantum state of all fields that are present. In this vein, we have considered small tensor and probe scalar fluctuations, to assess the quantum state they obtain due to the wormhole evolution, and which can be considered as their initial state for the subsequent evolution in Lorentzian time.

The results we find are intriguing: we show that the wormholes amplify perturbation modes in such a way that they behave very nearly like the Bunch-Davies ground state in inflationary cosmology. (This can be understood as being largely due to the fact that the scalar field that supports the wormhole solution interpolates to positive values of its potential.) However, the wormholes do not lead to the exact BD vacuum -- rather, they generate small deviations which scale inversely with the square of the wavenumber. Physically, this is highly reasonable: the perturbations with the smallest wavenumber have a wavelength that is comparable to the radius of the wormhole, and thus they are very sensitive to the wormhole geometry. By contrast, perturbations with a very short wavelength feel the wormhole geometry much less and consequently remain closer to the standard vacuum state.

The deviations that we have just mentioned occur equally for tensor and scalar fluctuations. When evolved forward in Lorentzian time, to super-horizon scales during the inflationary phase that follows the nucleation of the baby universe, these deviations shift the angular power spectrum of the now amplified perturbations, and moreover introduce small oscillations, again with an envelope that decays inversely with the square of the wavenumber. Thus the wormholes lead to a very distinctive signal in the spectrum of all perturbative modes. The amplitude of the deviations depends quite strongly on the wormhole charge, and can reach several percent of the power spectrum when the charge is large.

An obvious question is whether such deviations are observable, i.e. whether we could determine, by looking at the CMB, whether our universe arose via a wormhole geometry. This is unfortunately too early to tell: since the deviations occur principally on the largest scales, their observability depends crucially on the length of the inflationary phase. If inflation lasted for a long time, much longer than the minimum required to explain the spatial flatness of the current state of the universe, then the small-wavenumber modes are still a long way beyond our current horizon, and will only become observable in the distant future. However, if inflation was more minimal, then such deviations might already be present on the CMB sky. A precise comparison with observations would require a knowledge of the inflationary potential, and this fact prevents us from making any detailed predictions at the current time. Let us just note that there are indeed deviations from the standard BD vacuum that are seen on large scales in the CMB (the so-called low-multipole anomalies, see e.g. \cite{Copi:2010na}), though there is the added caveat that all such large-scale modes suffer from the inherent randomness induced by cosmic variance (very few such long-wavelength modes fit on the sky, so that their expected amplitude is statistically more uncertain).

Also, we should keep in mind that we only considered probe scalar modes in the present work. An important subject for future investigation is the study of fluctuations in the scalar field that supports the wormhole solutions. This may prove challenging, due to a technical complication known as the negative-mode problem \cite{Lavrelashvili:1985vn,Tanaka:1992zw,Lavrelashvili:1998dt,Tanaka:1999pj,Khvedelidze:2000cp,Lavrelashvili:1999sr,Gratton:2000fj,Lavrelashvili:2006cv,Dunne:2006bt,Battarra:2012vu,Lee:2014uza,Koehn:2015hga,Bramberger:2019mkv,Jinno:2020zzs}. Specifically, the kinetic factor in the quadratic action of physical perturbations depends on the background scale factor and scalar field profile, as well as the wavenumber of the perturbations, and it becomes negative along the wormhole, at least for homogeneous perturbations. When it becomes negative, formally this leads to infinitely many unphysical negative modes and renders the corresponding Sturm-Liouville problem intractable (though the suspicion is that this difficulty might be due to an unfortunate choice of perturbation variable). With the addition of an axion field, the negative mode problem becomes even more singular \cite{Jonas:2023qle}. However, we note that in false vacuum decay with gravity, this problem affects only the homogeneous sector (see the kinetic factor expression in, e.g., \cite{Khvedelidze:2000cp}). Therefore, there is some hope that, based on the formalism developed in \cite{Jonas:2023qle,Hertog:2024nys}, one can successfully study scalar perturbations for harmonics with $n\geq 2$. We leave this for future work.

An important corollary of our work arises when looking at wormholes with small (axionic or magnetic) charges. Our results show that the deviations from the BD vacuum are proportional to a positive power of the charges. This means that for small charges, they are suppressed and in fact one would expect an exact BD state to be recovered in the limit of vanishing charge. But this is precisely the limit in which we argued in previous work \cite{Lavrelashvili:2026zsw,Lavrelashvili:2026jcw} that the wormhole geometry splits into two separate pieces, one being the asymptotic background and the other becoming a disconnected dS (no-boundary) instanton. Thus we find further evidence that this topological transition is indeed smooth and that the inclusion of perturbations does not invalidate the arguments presented in our earlier works, as perturbations do not blow up in the stem region when it shrinks away.

Based on the just-described small-charge limit of the wineglass wormholes and their connection to dS instantons, a new approach to setting boundary conditions in quantum cosmology has recently been suggested in \cite{Yamada:2026lzg}. Given our results for small deviations from Bunch-Davies perturbations, it will be interesting to see if in this new approach the perturbations also acquire a distinctive signature. This is a further topic for future research.

\section*{Acknowledgements}
The authors thank the Albert-Einstein-Institute for kind hospitality while this work was conducted. The work of G.L. is supported in part by the Shota Rustaveli National Science Foundation of Georgia with Grant No. FR-24-901.

\bibliographystyle{utphys}
\bibliography{biblio}

@article{Coleman:1980aw,
    author = "Coleman, Sidney R. and De Luccia, Frank",
    title = "{Gravitational Effects on and of Vacuum Decay}",
    reportNumber = "SLAC-PUB-2463",
    doi = "10.1103/PhysRevD.21.3305",
    journal = "Phys. Rev. D",
    volume = "21",
    pages = "3305",
    year = "1980"
}

@article{Lavrelashvili:1985vn,
    author = "Lavrelashvili, George  and Rubakov, V. A. and Tinyakov, P. G.",
    title = "{Tunneling transitions with gravitation: breaking of the quasiclassical approximation}",
    doi = "10.1016/0370-2693(85)90761-0",
    journal = "Phys. Lett. B",
    volume = "161",
    pages = "280--284",
    year = "1985"
}

@article{Lavrelashvili:1988un,
    author = "Lavrelashvili, G. and Rubakov, V. A. and Tinyakov, P. G.",
    title = "{Loss of Quantum Coherence Due to Topological Changes: A Toy Model}",
    doi = "10.1142/S0217732388001483",
    journal = "Mod. Phys. Lett. A",
    volume = "3",
    pages = "1231--1242",
    year = "1988"
}

@article{Tanaka:1992zw,
    author = "Tanaka, Takahiro and Sasaki, Misao",
    title = "{False vacuum decay with gravity: Negative mode problem}",
    reportNumber = "KUNS-1135",
    doi = "10.1143/PTP.88.503",
    journal = "Prog. Theor. Phys.",
    volume = "88",
    pages = "503--528",
    year = "1992"
}

@article{Lavrelashvili:1998dt,
    author = "Lavrelashvili, George",
    title = "{On the quadratic action of the Hawking-Turok instanton}",
    eprint = "gr-qc/9804056",
    archivePrefix = "arXiv",
    reportNumber = "BUTP-98-12",
    doi = "10.1103/PhysRevD.58.063505",
    journal = "Phys. Rev. D",
    volume = "58",
    pages = "063505",
    year = "1998"
}

@article{Tanaka:1999pj,
    author = "Tanaka, Takahiro",
    title = "{The No - negative mode theorem in false vacuum decay with gravity}",
    eprint = "gr-qc/9901082",
    archivePrefix = "arXiv",
    reportNumber = "OU-TAP-92",
    doi = "10.1016/S0550-3213(99)00369-7",
    journal = "Nucl. Phys. B",
    volume = "556",
    pages = "373--396",
    year = "1999"
}

@article{Khvedelidze:2000cp,
    author = "Khvedelidze, Arsen and Lavrelashvili, George and Tanaka, Takahiro",
    title = "{On cosmological perturbations in closed FRW model with scalar field and false vacuum decay}",
    eprint = "gr-qc/0001041",
    archivePrefix = "arXiv",
    doi = "10.1103/PhysRevD.62.083501",
    journal = "Phys. Rev. D",
    volume = "62",
    pages = "083501",
    year = "2000"
}

@article{Lavrelashvili:1999sr,
    author = "Lavrelashvili, George",
    editor = "de Alfaro, Vittorio and Nelson, J. E. and Cadoni, M. and Cavaglia, M. and Filippov, A.",
    title = "{Negative mode problem in false vacuum decay with gravity}",
    eprint = "gr-qc/0004025",
    archivePrefix = "arXiv",
    doi = "10.1016/S0920-5632(00)00756-8",
    journal = "Nucl. Phys. B Proc. Suppl.",
    volume = "88",
    pages = "75--82",
    year = "2000"
}

@article{Gratton:2000fj,
    author = "Gratton, Steven and Turok, Neil",
    title = "{Homogeneous modes of cosmological instantons}",
    eprint = "hep-th/0008235",
    archivePrefix = "arXiv",
    doi = "10.1103/PhysRevD.63.123514",
    journal = "Phys. Rev. D",
    volume = "63",
    pages = "123514",
    year = "2001"
}

@article{Lavrelashvili:2006cv,
    author = "Lavrelashvili, George",
    title = "{The Number of negative modes of the oscillating bounces}",
    eprint = "gr-qc/0602039",
    archivePrefix = "arXiv",
    doi = "10.1103/PhysRevD.73.083513",
    journal = "Phys. Rev. D",
    volume = "73",
    pages = "083513",
    year = "2006"
}

@article{Dunne:2006bt,
    author = "Dunne, Gerald V. and Wang, Qing-hai",
    title = "{Fluctuations about Cosmological Instantons}",
    eprint = "hep-th/0605176",
    archivePrefix = "arXiv",
    doi = "10.1103/PhysRevD.74.024018",
    journal = "Phys. Rev. D",
    volume = "74",
    pages = "024018",
    year = "2006"
}

@article{Battarra:2012vu,
    author = "Battarra, Lorenzo and Lavrelashvili, George and Lehners, Jean-Luc",
    title = "{Negative Modes of Oscillating Instantons}",
    eprint = "1208.2182",
    archivePrefix = "arXiv",
    primaryClass = "hep-th",
    doi = "10.1103/PhysRevD.86.124001",
    journal = "Phys. Rev. D",
    volume = "86",
    pages = "124001",
    year = "2012"
}

@article{Lee:2014uza,
    author = "Lee, Hakjoon and Weinberg, Erick J.",
    title = "{Negative modes of Coleman-De Luccia bounces}",
    eprint = "1408.6547",
    archivePrefix = "arXiv",
    primaryClass = "hep-th",
    doi = "10.1103/PhysRevD.90.124002",
    journal = "Phys. Rev. D",
    volume = "90",
    number = "12",
    pages = "124002",
    year = "2014"
}

@article{Koehn:2015hga,
    author = "Koehn, Michael and Lavrelashvili, George and Lehners, Jean-Luc",
    title = "{Towards a Solution of the Negative Mode Problem in Quantum Tunnelling with Gravity}",
    eprint = "1504.04334",
    archivePrefix = "arXiv",
    primaryClass = "hep-th",
    doi = "10.1103/PhysRevD.92.023506",
    journal = "Phys. Rev. D",
    volume = "92",
    number = "2",
    pages = "023506",
    year = "2015"
}

@article{Bramberger:2019mkv,
    author = "Bramberger, Sebastian F. and Chitishvili, Mariam and Lavrelashvili, George",
    title = "{Aspects of the negative mode problem in quantum tunneling with gravity}",
    eprint = "1906.07033",
    archivePrefix = "arXiv",
    primaryClass = "gr-qc",
    doi = "10.1103/PhysRevD.100.125006",
    journal = "Phys. Rev. D",
    volume = "100",
    number = "12",
    pages = "125006",
    year = "2019"
}

@article{Jinno:2020zzs,
    author = "Jinno, Ryusuke and Sato, Ryosuke",
    title = "{Negative mode problem of false vacuum decay revisited}",
    eprint = "2010.04462",
    archivePrefix = "arXiv",
    primaryClass = "hep-th",
    reportNumber = "DESY-20-171, DESY 20-171",
    doi = "10.1103/PhysRevD.104.096009",
    journal = "Phys. Rev. D",
    volume = "104",
    number = "9",
    pages = "096009",
    year = "2021"
}

@article{Lehners:2023yrj,
    author = "Lehners, Jean-Luc",
    title = "{Review of the no-boundary wave function}",
    eprint = "2303.08802",
    archivePrefix = "arXiv",
    primaryClass = "hep-th",
    doi = "10.1016/j.physrep.2023.06.002",
    journal = "Phys. Rept.",
    volume = "1022",
    pages = "1--82",
    year = "2023"
}

@article{Jonas:2023ipa,
    author = "Jonas, Caroline and Lavrelashvili, George and Lehners, Jean-Luc",
    title = "{Zoo of axionic wormholes}",
    doi = "10.1103/PhysRevD.108.066012",
    journal = "Phys. Rev. D",
    volume = "108",
    number = "6",
    pages = "066012",
    year = "2023"
}

@article{Jonas:2023qle,
    author = "Jonas, Caroline and Lavrelashvili, George and Lehners, Jean-Luc",
    title = "{Stability of axion-dilaton wormholes}",
    eprint = "2312.08971",
    archivePrefix = "arXiv",
    primaryClass = "hep-th",
    doi = "10.1103/PhysRevD.109.086022",
    journal = "Phys. Rev. D",
    volume = "109",
    number = "8",
    pages = "086022",
    year = "2024"
}

@article{Hertog:2024nys,
    author = "Hertog, T. and Maenaut, S. and Missoni, B. and Tielemans, R. and Van Riet, T.",
    title = "{Stability of axion-saxion wormholes}",
    eprint = "2405.02072",
    archivePrefix = "arXiv",
    primaryClass = "hep-th",
    doi = "10.1007/JHEP11(2024)151",
    journal = "JHEP",
    volume = "11",
    pages = "151",
    year = "2024"
}

@article{Betzios:2024oli,
    author = "Betzios, Panos and Papadoulaki, Olga",
    title = "{Inflationary Cosmology from Anti-de Sitter Wormholes}",
    eprint = "2403.17046",
    archivePrefix = "arXiv",
    primaryClass = "hep-th",
    doi = "10.1103/PhysRevLett.133.021501",
    journal = "Phys. Rev. Lett.",
    volume = "133",
    number = "2",
    pages = "021501",
    year = "2024"
}

@article{Betzios:2024zhf,
    author = "Betzios, Panos and Gialamas, Ioannis D. and Papadoulaki, Olga",
    title = "{Magnetic anti{\textendash}de Sitter wormholes as seeds for Higgs inflation}",
    eprint = "2412.03639",
    archivePrefix = "arXiv",
    primaryClass = "hep-th",
    doi = "10.1103/9w85-fyhs",
    journal = "Phys. Rev. D",
    volume = "111",
    number = "12",
    pages = "123542",
    year = "2025"
}

@article{Lan:2024gnv,
    author = "Lan, Qing-Yu and Piao, Yun-Song",
    title = "{Prepare inflationary universe via the Euclidean charged wormhole}",
    eprint = "2411.13844",
    archivePrefix = "arXiv",
    primaryClass = "gr-qc",
    month = "11",
    year = "2024"
}

@article{Betzios:2026rbv,
    author = "Betzios, Panos and Ghiringhelli, Paul and Gialamas, Ioannis D. and Papadoulaki, Olga",
    title = "{A Menagerie of Wormholes and Cosmologies in the Gravitational Path Integral}",
    eprint = "2602.23432",
    archivePrefix = "arXiv",
    primaryClass = "hep-th",
    month = "2",
    year = "2026"
}

@article{Lavrelashvili:2026zsw,
    author = "Lavrelashvili, George and Lehners, Jean-Luc",
    title = "{Nucleating an Inflationary Universe: Euclidean Wormholes and their No-Boundary Limit}",
    eprint = "2603.11003",
    archivePrefix = "arXiv",
    primaryClass = "hep-th",
    month = "3",
    year = "2026"
}

@article{Planck:2018jri,
    author = "Akrami, Y. and others",
    collaboration = "Planck",
    title = "{Planck 2018 results. X. Constraints on inflation}",
    eprint = "1807.06211",
    archivePrefix = "arXiv",
    primaryClass = "astro-ph.CO",
    doi = "10.1051/0004-6361/201833887",
    journal = "Astron. Astrophys.",
    volume = "641",
    pages = "A10",
    year = "2020"
}

@article{Lavrelashvili:2026jcw,
    author = "Lavrelashvili, George and Lehners, Jean-Luc",
    title = "{Birth of Inflationary Universes via Wineglass Wormholes and their No-Boundary Relatives}",
    eprint = "2605.10548",
    archivePrefix = "arXiv",
    primaryClass = "hep-th",
    month = "5",
    year = "2026"
}

@article{Betzios:2026fgf,
    author = "Betzios, Panos and Ghiringhelli, Paul and Gialamas, Ioannis D. and Papadoulaki, Olga",
    title = "{Before the Bang: Wormholes at the Dawn of the Universe}",
    eprint = "2605.13777",
    archivePrefix = "arXiv",
    primaryClass = "hep-th",
    month = "5",
    year = "2026"
}

@article{Khan:2026doo,
    author = "Khan, Imtiaz and Pirzada and Mustafa, G. and Atamurotov, Farruh and Yuan, Chengxun",
    title = "{Long Inflation Screens Euclidean-Wormhole Initial States}",
    eprint = "2605.04166",
    archivePrefix = "arXiv",
    primaryClass = "hep-ph",
    month = "5",
    year = "2026"
}

@article{Yamada:2026lzg,
    author = "Yamada, Masaki",
    title = "{A Small-Throat Boundary Condition for the Tunneling Wave Function of the Universe}",
    eprint = "2607.08815",
    archivePrefix = "arXiv",
    primaryClass = "gr-qc",
    reportNumber = "TU-1311",
    month = "7",
    year = "2026"
}

@article{Bunch:1978yq,
    author = "Bunch, T. S. and Davies, P. C. W.",
    title = "{Quantum Field Theory in de Sitter Space: Renormalization by Point Splitting}",
    doi = "10.1098/rspa.1978.0060",
    journal = "Proc. Roy. Soc. Lond. A",
    volume = "360",
    pages = "117--134",
    year = "1978"
}

@article{Gratton:2001gw,
    author = "Gratton, Steven and Lewis, Antony and Turok, Neil",
    title = "{Closed universes from cosmological instantons}",
    eprint = "astro-ph/0111012",
    archivePrefix = "arXiv",
    doi = "10.1103/PhysRevD.65.043513",
    journal = "Phys. Rev. D",
    volume = "65",
    pages = "043513",
    year = "2002"
}

@article{Feldbrugge:2017fcc,
    author = "Feldbrugge, Job and Lehners, Jean-Luc and Turok, Neil",
    title = "{No smooth beginning for spacetime}",
    eprint = "1705.00192",
    archivePrefix = "arXiv",
    primaryClass = "hep-th",
    doi = "10.1103/PhysRevLett.119.171301",
    journal = "Phys. Rev. Lett.",
    volume = "119",
    number = "17",
    pages = "171301",
    year = "2017"
}

@article{Lehners:2021jmv,
    author = "Lehners, Jean-Luc",
    title = "{Wave function of simple universes analytically continued from negative to positive potentials}",
    eprint = "2105.12075",
    archivePrefix = "arXiv",
    primaryClass = "hep-th",
    doi = "10.1103/PhysRevD.104.063527",
    journal = "Phys. Rev. D",
    volume = "104",
    number = "6",
    pages = "063527",
    year = "2021"
}

@article{Gerlach:1978gy,
    author = "Gerlach, U. H. and Sengupta, U. K.",
    title = "{Homogeneous Collapsing Star: Tensor and Vector Harmonics for Matter and Field Asymmetries}",
    doi = "10.1103/PhysRevD.18.1773",
    journal = "Phys. Rev. D",
    volume = "18",
    pages = "1773--1784",
    year = "1978"
}

@article{Copi:2010na,
    author = "Copi, Craig J. and Huterer, Dragan and Schwarz, Dominik J. and Starkman, Glenn D.",
    title = "{Large angle anomalies in the CMB}",
    eprint = "1004.5602",
    archivePrefix = "arXiv",
    primaryClass = "astro-ph.CO",
    doi = "10.1155/2010/847541",
    journal = "Adv. Astron.",
    volume = "2010",
    pages = "847541",
    year = "2010"
}

\end{document}